\documentclass[12pt]{article}
\usepackage[dvips]{graphicx}
\usepackage{pdproc}

  \makeatletter 
  \def\@cite#1{[#1]} 
  \makeatother    
\begin{document}

\renewcommand{\thefootnote}{\alph{footnote}}

\title{
 Non-BPS walls and their stability in 5d SUSY theory\footnote{
 Talk given by M.E.
 at 12th International Conference 
on Supersymmetry and Unification of Fundamental Interactions
June 17-23, 2004, Epochal Tsukuba, Tsukuba, Japan
}}

\author{ M.Eto~$\!\!^{1}\!\!$, N.Maru~$\!\!^{2}\!\!$ and N.Sakai~$\!\!^{1}\!\!$}

\address{ 
$^{1}$Department of Physics, Tokyo Institute of Technology \\
Tokyo, 152-8551, Japan\\
$^{2}$Theoretical Physics Laboratory, RIKEN \\ 
 Saitama 351-0198, JAPAN 
%%%%% You may comment out the e-mail address line below.  
%\\ {\rm E-mail: 
%meto@th.phys.titech.ac.jp (M.Eto),
%maru@postman.riken.go.jp (N.Maru),
%nsakai@th.phys.titech.ac.jp (N.Sakai)}
}

\abstract{
We review recent work (hep-th/0404114)
on non-BPS domain walls in $5d$ SUSY theory.
An exact solution of non-BPS multi-walls 
is found in supersymmetric massive 
$T^\star(\bf{CP}^1)$ model 
in five dimensions. 
The non-BPS multi-wall solution is found to have 
no tachyon. 
The ${\cal N}=1$ supersymmetry preserved 
on the four-dimensional world volume of one wall 
is broken by the coexistence of the other wall. 
The supersymmetry breaking is exponentially 
suppressed as the distance between the walls increases.  
}

\normalsize\baselineskip=15pt
\ \\

%\section{Introduction}
Brane-world scenario with 
extra dimensions have attracted much attention 
in recent years. 
It has also been useful to implement supersymmetry 
(SUSY) to obtain solitons such as walls 
where particles should be localized. 
SUSY has been most useful to obtain 
realistic unified theories, but 
one of their least understood problems 
is the origin of the SUSY breaking. 
It has been found that the coexistence of 1/2 BPS walls 
can break SUSY completely, leading to a possible 
origin of the SUSY breaking\cite{Maru:2000sx}. 
On the other hand, the stability of 
such non-BPS configurations 
are no longer guaranteed by SUSY. 
If we introduce topological quantum numbers, we can have a 
stable non-BPS walls in four-dimensional theories
\cite{Maru:2001gf,SS}. 
However, models in five dimensions are needed 
to obtain realistic models for brane-world. 
The BPS wall solutions of hypermultiplets 
have been obtained for the simplest of 
such nonlinear sigma models, a massive 
$T^\star(\bf{CP}^1)$ SUSY and SUGRA model\cite{Arai:2002xa,AFNS,EFNS}. 
However, no non-BPS multi-wall solutions have been 
obtained so far. 
The purpose of our paper is to present 
non-BPS multi-wall solutions in a SUSY theory in five 
dimensions and to discuss their stability. 

%\section{Non-BPS walls in five dimesional SUSY}

In order to construct stable domain walls,
the models should have discrete SUSY vacua.
In five dimensional SUSY theory, 
this can be achieved either 
by a SUSY gauge theories interacting with hypermultiplets, 
or by nonlinear sigma models of hypermultiplets. 
As a gauge theory, one can take a SUSY $U(N_c)$ gauged 
theory with $N_f>N_c$ flavors of 
hypermultiplets in the fundamental representation. 
If the hypermultiplet  masses are nondegenerate 
and the $U(1)$ factor group of $U(N_c)$ has the 
Fayet-Iliopoulos (FI) terms, the model exhibits 
discrete SUSY vacua.

For simplicity we will consider the SUSY $U(1)$ gauge 
theory with $N_f (\ge 2)$ hypermultiplets.
The scalar potential is given by
%bosonic Lagrangian is given by 
\begin{eqnarray}
%{\cal L} &=& 
%- \frac{1}{4e^2} F^{MN}F_{MN} - \frac{1}{2e^2} 
%\partial_M\Sigma\partial^M\Sigma 
%- \sum_{A=1}^{N_f}{\cal D}^MH_{Ai}^* {\cal D}_M H_A^i - V,\\
V &=& \frac{e^2}{2}\sum_{a=1}^3
\left( -2\xi\delta_3^a + 
\sum_{A=1}^{N_f} H^*_{Ai}(\sigma^a)^i{}_jH_A^j\right)^2
+ \sum_{A=1}^{N_f}\left(\Sigma + \mu_A\right)^2 H_{Ai}^*H_A^i, 
\end{eqnarray}
where $\mu_A(\neq\mu_B)$ is the mass of the $A$-th hypermultiplet.
The $SU(2)_R$ triplet FI parameters 
are chosen to lie in the third direction and is 
denoted as $\xi$. 
The $A$-the vacuum is given by 
$
\Sigma = - \mu_A,\
H_B^2 = 0,\
|H_B^1|^2 = 2\xi\delta^A_B\ 
(B=1,2,\cdots,N_f)$.
The minimal model admitting such a %single 
BPS domain 
wall is the case of $N_f=2$, which will be considered 
from now on.
Even with this simple model, it is generally 
difficult to obtain exact wall solutions for the case 
of finite gauge coupling~\cite{KakimotoSakai}, 
\cite{Isozumi:2003rp}. 
It is sufficient to examine 
the case of infinite 
gauge coupling to study domain walls
because the vacuum configuration does not depend on
the gauge coupling.
As we let gauge coupling to infinity 
$e\rightarrow \infty$, 
the kinetic term of vector multiplet in the Lagrangian 
vanishes. 
At the same time, the scalar potential becomes infinitely 
steep and the hypermultiplets are constrained to be at 
the minimum.
The gauge field $A_M$ and the adjoint scalar field 
$\Sigma$ in the vector multiplet become Lagrange 
multiplier fields.
Then the model reduces to the 
$T^\star(\bf{CP}^1)$ non-linear sigma model.
The sigma model can be expressed by a inhomogeneous coordinates
$(R, \Omega, \Theta, \Phi)$ \cite{Eto:2004zc,Curtright:1979yz,Gibbons:1987pk}.
The bosonic part of the Lagrangian reads 
\begin{eqnarray}
{\cal L}_\infty &=&
\frac{1}{2\sqrt{R^2 + \xi^2}}\bigg[
- \partial^MR \partial_M R - (R^2 + \xi^2)\partial_M
\Theta\partial^M\Theta
- \left(R^2 + \xi^2\sin^2\Theta\right)\partial_M
\Phi\partial^M\Phi\nonumber\\
&&- R^2\partial_M\Omega\partial^M\Omega 
- 2R^2\cos\Theta\partial_M\Phi\partial^M\Omega
- \mu^2\left(R^2 + \xi^2\sin^2\Theta\right)\bigg].
\label{EH_lagrangian}
\end{eqnarray}
Since a common mass of hypermultiplets can be absorbed 
by a shift of vector multiplet scalar $\Sigma$, we 
set $\mu_1 = -\mu_2 = \frac{\mu}{2}$ here. 
This model admits two SUSY vacua at $R=0,\ \Theta=0,\pi$ which
correspond to the north and south pole of the base manifold $\bf{CP}^1$
respectively.
We exactly solve the field equation and find
non-BPS domain wall solutions which interpolate these vacua:
\begin{eqnarray}
\Theta(y;y_0,k) 
= {\rm am}\left((\mu/k)(y-y_0),k\right) 
+ \pi/2,\quad
\Phi = {\rm const.},\quad
R = \Omega = 0.
\label{nbps_cp1}
\end{eqnarray}
Besides $y_0$ representing the position of the wall,
this solution has one more parameter $k$ and it 
has periodicity\footnote{$K$ is the complete elliptic
integral of the first kind.} $2\pi L = 4k K(k)/\mu$.
In the case of $k>1$ the solution curve never reaches 
either vacuum at north and south poles, and oscillates 
in an interval between them. 
In the case of $k=1$ the solution corresponds to
the BPS or anti-BPS solution
which interpolates north pole and south pole once.
In the case of $k<1$ the solution passes through 
both vacua and becomes quasi-periodic. 
Similarly to the sine-Gordon case, this solution 
represents the BPS wall and the anti-BPS wall 
placed at $y_0$ and at $y_0+\pi L$, respectively. 
In the following we will concentrate on this quasi-periodic
solution $(k<1)$ only.

%\section{Stability}
To examine the stability of the exact non-BPS domain wall 
solution (\ref{nbps_cp1}) in the massive 
$T^\star(\bf{CP}^1)$ 
nonlinear sigma model in five dimensions, 
we first study small fluctuations around our background solution:
$
\Theta(x^m,y) 
= \Theta_0(y) + \theta(x^m,y),
\Phi(x^m,y) = \Phi_0 + \varphi(x^m,y), 
R(x^m,y) = r(x^m,y),
\Omega(x^m,y) = \omega(x^m,y). 
$
The part of the Lagrangian quadratic in the fluctuations 
is decomposed into a sum 
for each fields 
\begin{eqnarray}
{\cal L}^{(2)}_{\rm boson}
&=&
 \int dy\ 
\bigg[
\xi/2\left\{- \partial^M\theta\partial_M\theta 
- \mu^2\cos2\Theta_0 \theta^2
- \sin^2\Theta_0 
\partial^M\varphi\partial_M\varphi
\right\}\nonumber\\
&+& \left\{  
-\partial^Mr \partial_Mr 
- \left(\mu^2 + \Theta_0'{^2}/2
- \mu^2\sin^2\Theta_0\right)r^2
\right\}/2\xi
\bigg]
.
\end{eqnarray}
Let us note that fluctuation of $\Omega$ disappears from 
the quadratic Lagrangian completely. 
For the fluctuations $r,\theta$ and $\tilde\varphi \equiv \sin\Theta_0 \varphi$,
we can define Shr\"odinger-type equations for mode functions
$\psi^{(n)}_X$ with mass squared $m^2_{X,n}$ of effective fields
on world volume as eigenvalues $(X=r,\theta,\tilde\varphi)$:
\begin{eqnarray}
\left[-\partial^2/\partial y^2 + {\cal V}_X(y)\right]\psi^{(n)}_X(y)
= m^2_{X,n}\psi_X^{(n)},
\end{eqnarray}
where the Schr\"odinger potentials are given by
\begin{eqnarray}
{\cal V}_r(y) &=&
\mu^2 + \Theta'_0{^2}/2 - (\mu^2/2)\sin^2\Theta_0 
= \mu^2(1 + k^2)/2k^2,\\
{\cal V}_\theta(y) &=& 
\mu^2 \cos 2\Theta_0 = 
\mu^2\left\{ 2{\rm sn}^2\left((\mu/k)(y-y_0),k\right) - 1\right\},\\
{\cal V}_{\varphi}(y) 
&=& \Theta_0''\tan\Theta_0 - (\Theta_0')^2
=
\mu^2
\left\{2 {\rm sn}^2\left((\mu/k)(y-y_0),k\right) 
- 1/k^2 
\right\}
.
\end{eqnarray}
If the tachyonic mode which has negative eigenvalue of the 
mode equation exists, it is indication that 
our non-BPS background configuration is unstable against small
perturbations. Clearly, there are no tachyonic mode in spectra of 
the fluctuations $r$ since corresponding Schr\"odinger potential
is positive constant.
We exactly solve the mode equation of $\theta$ for the first several 
lightest modes:
\begin{eqnarray}
m_{\theta,0}^2 = 0,\quad 
\psi_\theta^{(0)} = N_\theta^{(0)}{\rm dn}
\left((\mu/k)(y-y_0),k\right),
\\
m_{\theta,1}^2 = \mu^2(1-k^2)/k^2,\quad 
\psi_\theta^{(1)} 
= N_\theta^{(1)}{\rm cn}\left((\mu/k)(y-y_0),k\right),
\\
m_{\theta,2}^2 = \mu^2/k^2,\quad 
\psi_\theta^{(2)} 
= N_\theta^{(2)}{\rm sn}\left((\mu/k)(y-y_0),k\right)
.
\end{eqnarray}
We can also exactly solve the mode equation for $\tilde\varphi$
because the Schr\"odinger potential ${\cal V}_{\tilde\varphi}$
is identical to that for $\theta$ except a constant shift:
${\cal V}_{\theta} - {\cal V}_{\tilde\varphi} = \mu^2(1-k^2)/k^2$.
Therefore the same eigenfunctions as %those for 
$\theta$ solve the eigenvalue problem for $%\tilde 
\varphi$ 
and the corresponding mass squared are 
shifted accordingly.
%\begin{eqnarray}
%m_{\varphi,-1}^2 = - \frac{1 - k^2}{k^2}\mu^2,
%\quad \psi_{\tilde\varphi}^{(-1)} 
%= N^{(-1)}_{\tilde\varphi}{\rm dn}
%\left(\frac{\mu}{k}(y-y_0),k\right),
%\\
%\quad m_{\varphi,0}^2 = 0,\quad 
%\psi_{\tilde\varphi}^{(0)} 
%= N^{(0)}_{\tilde\varphi}{\rm cn}\left(\frac{\mu}{k}(y-y_0),k\right),
%\quad\\
%\quad m_{\varphi,1}^2 = \mu^2,\quad 
%\psi_{\tilde\varphi}^{(1)} 
%= N^{(1)}_{\tilde\varphi}{\rm sn}\left(\frac{\mu}{k}(y-y_0),k\right).\quad
%\end{eqnarray}
In contrast to the case of  $\theta$, 
the solution
at first sight appears to indicate instability 
of the background solution, contrary to our expectations. 
However, the tachyonic mode $\psi_{\tilde\varphi}^{(-1)}$ is 
not normalizable and unphysical. 
Thus we conclude that there is no tachyonic modes in the spectra
around our non-BPS background solution.

Our non-BPS solution is very similar to the non-BPS stable solution which
was found in the four dimensional sine-Gordon model\cite{Maru:2000sx}.
However, our model has trivial first homotopy $\pi_1(\bf{CP}^1)=0$,
so there is no reason which supports the stability of our background
configuration unlike the sine-Gordon model\cite{Maru:2000sx}.
To verify instability under {\it large} deformations, we also examine
a continuous deformation which makes the wall path shrinking to a
point on $S^2$, in the spirit of variational approach.
For simplicity, we consider a path on $S^2$ which 
%start at $\Theta=0$, 
cuts off our non-BPS solution at $\Theta$, 
turns around a circle of $\Phi$ rotating by $\pi$ with 
the constant $\Theta$, and going to back through 
our non-BPS solution reflected at $\Theta=\pi$
\begin{eqnarray}
\Theta_1(u;u_\star) = \left\{
\begin{array}{l}
{\rm am}(u,k) + \frac{\pi}{2},\\
\Theta_\star \equiv {\rm am}(u_\star,k) + \frac{\pi}{2},\\
\frac{3\pi}{2} - {\rm am}(u,k)
\end{array}
\right.
\Phi_1(u;u_\star) = \left\{
\begin{array}{ll}
0 &   -K < u \le u_\star,\\
\frac{\pi}{2}\frac{u-K}{K-u_\star} + \frac{\pi}{2}
&  u_\star < u \le 2K - u_\star,\\
\pi &  2K - u_\star < u \le 3K,
\end{array}
\right.
\end{eqnarray}
where we have defined a variable 
$u \equiv {\mu \over k}(y-y_0)$
and $u_\star$ denotes the position in extra dimension 
corresponding to the value $\Theta=\Theta_\star$. 
The energy of the above trial function is given by $(\ell \equiv K -u_\star)$
\begin{eqnarray}
{\cal E}(\ell) = \frac{\mu\xi}{k}\left[
E(K\!-\ell)\! -\! E(-K) + (k^2 - 1)(2K-\ell)
+ \left(\frac{\pi^2}{4\ell} + k^2\ell\right)
{\rm cn}^2(K\!-\ell,k)
\right]. 
\end{eqnarray}
We observe that the energy of the path of 
the continuous deformation has a local minimum at 
our non-BPS solution, in accordance with our result 
of no tachyon under small fluctuations. 
It then shows a maximum before reaching to the 
absolute minimum at the true vacuum. 
We regard this result as an evidence for the 
instability under large fluctuations. 

Let us close by mentioning spontaneously SUSY breaking by
the co-existing BPS and anti-BPS walls.
From the brane world viewpoint, 
it is interesting to study how SUSY breaking 
effects are generated 
on a wall by the existence of the other walls. 
Once SUSY is broken by the vacuum expectation values 
of auxiliary fields of some supermultiplets 
in the hidden brane, SUSY breaking effects 
are mediated to the visible brane 
by bulk fields interacting with both branes. 
Then, soft SUSY breaking terms of MSSM 
fields on the visible brane 
are generated. 
In this framework, various fields have to 
be added on the hidden brane 
and/or in the bulk by hand to break SUSY and 
to transmit the SUSY breaking effects 
to our world. 
On the other hand, 
we have no need to add extra fields mentioned above 
since the non-BPS configuration itself breaks SUSY and 
the fields forming the non-BPS wall 
are responsible for SUSY breaking and 
its transmission to our world.  
The mass splitting 
$\Delta m^2 \equiv m^2_{{\rm boson}} - m^2_{{\rm fermion}}$ 
is simply given by 
$\Delta m^2 = m^2_{\theta,1}=\frac{1-k^2}{k^2}\mu^2$. 
This mass splitting can be related to the distance between 
the walls. 
In the limit $k \to 1$, we obtain 
$
\Delta m^2 = \mu^2 \frac{e^{-\pi \mu L}}{1-e^{-\pi \mu L}} 
\simeq \mu^2 e^{-\pi \mu L}. 
$
The mass splitting is exponentially suppressed 
as a function of 
the distance $\pi L$ between walls. 
This result is also phenomenologically fascinating 
in that the low SUSY breaking scale can be 
naturally generated 
from the five-dimensional Planck scale 
$\mu \sim {\cal O}(M_5)$ 
without an extreme fine-tuning of parameters. 

\bibliographystyle{plain}

\end{document}